\documentclass[twocolumn]{svjour3}         % twocolumn
\smartqed  % flush right qed marks, e.g. at end of proof
\usepackage{graphicx}
\usepackage{marvosym}

\usepackage{multirow}

\begin{document}

%\title{Local strain and stress measurements in soft grain piles
\title{Mapping forces in 3D elastic assembly of grains.
}


\author{Mohammad Saadatfar \and Adrian P. Sheppard\and Tim J. Senden  \and \\ Alexandre J. Kabla}


\institute{M. Saadatfar \and T. J. Senden\and A. P. Sheppard \at 
Research School of Physics and Engineering, The Australian National University, Canberra ACT 0200 Australia
\and
A. J. Kabla \at 
Department of Engineering, University of Cambridge, Trumpington street, CB2 1PZ, Cambridge, UK\\
%\email{} 
% (\Letter)
}

\maketitle

\begin{abstract}
Our understanding of the elasticity and rheology of disordered materials, such as granular piles, 
foams, emulsions or dense suspensions
relies on improving experimental tools to characterize their behaviour at the
particle scale. While 2D observations are now routinely carried out in
laboratories, 3D measurements remain a challenge. In this paper, we use a simple model system,  
a packing of soft elastic spheres, to illustrate the capability of X-ray
microtomography to characterise the internal structure and local behaviour of granular systems. Image analysis
techniques can resolve grain positions, shapes and contact areas; this is used to 
investigate the material's microstructure and its evolution upon strain. In
addition to morphological measurements, we develop a technique to quantify contact forces and estimate the internal stress tensor. 
As will be illustrated in this paper, this opens the door to a broad array of static and dynamical measurements in 3D disordered systems.
\end{abstract}

%\section{Introduction}
\label{intro}

Despite the practical importance of macroscopic disordered materials such as soil, foams and emulsions the 
rules that govern their mechanical behaviour remain poorly understood. In the case of granular materials, 
these studies are essential to the advancement of related industrial processes and also to the 
prediction of often catastrophic geological phenomena (avalanches of slurries, earthquakes). 
Although empirical constitutive equations are nowadays able to partly model the response of these 
systems, a unified multiscale framework does not exist yet. Over the past 20 years, foams and granular 
systems have emerged as a model system for low temperature glasses \cite{Langer:00}. Many theories 
and experiments have been designed to address the relationship between the internal structure and macroscopic 
behaviour of such systems. These theories and models often use approaches derived from statistical 
physics \cite{Zhang:05a,OHern:01}. Although the behaviour of foams seems today rather well 
understood \cite{Weaire2010}, the complexity of granular matters behaviour at the microscopic 
scale has prevented a proper physical model of granular systems.
A large part of the problem arises from the strongly nonlinear contact law between rigid bodies coupled 
with the dynamical nature of the contact network. In this context, the spatial distribution and temporal 
evolution of the force network has become a highly sought after quantity that expands from micro 
scale (grain-grain contacts) to macro scale (granular assembly).

Early simulations have provided important insights on the internal force distributions 
of a granular pile \cite{Radjai:96,Coppersmith:96,Makse:00a}. Two-dimensional (2D) experiments 
managed to keep up with the pace of advancement in simulations \cite{Drescher:72,Howell:99,Kolb:04}, 3D 
experiments however, have been limited to measurements of the distribution of forces only at the boundaries of the
containers \cite{Brockbank:97,Mueth:98,Blair:01,Corwin:05}. These experimental methods could not 
have access to the spatial arrangement of the contact force network in the bulk of granular 
assemply. Moreover they were unable to determine structural features such as force chains and 
arching which have been postulated as the signature of jamming \cite{Lois:07}.

In recent years, a range of tools have been developed to apprehend the 3D
nature of bulk properties. X-ray or MRI techniques have allowed to study dynamic properties 
of granular systems such as flow profiles and shear banding, at a mesoscopic 
scale \cite{Mantle:01,Alex:05,Alex:09,Dijksman:10}. 3D reconstructions of compressed 
emulsion systems using confocal microscopy have provided the first measurements of the bulk 
force distributions \cite{Brujic:03,Bruji:05} by estimating the contact force from 
contact geometry. A similar technique has also been used to characterise spatial 
correlations of forces inside 3D piles of frictionless liquid droplets \cite{Zhou06}. 
Although these observations are very valuable, the systems used are significantly different 
from real granular pile which has very different contact laws and frictional properties.

X-ray computed tomography has been used for static characterisation of large packings 
of spherical grains \cite{Silb:02,Aste:05,saadatfarthesis}. In this study, we apply this 
technique to a model system made of soft millimetric elastic beads. In addition to the 
measurement of traditional structural quantities (packing fraction, coordination number etc.), 
the contact force is measured from the contact area between grains which can be integrated at a 
mesoscopic scale to estimate the local stress tensor in the pile. 
We then apply a series of external forces to our model system and investigate its response to these 
controlled loadings. This approach delivers results consistent with the existing 
literature and is promising as a generic tool to study the local, non-linear mechanics of granular 
assemblies.

\section{Experiment and methodology}

\subsection{Experimental setup}

\begin{figure}
\centering
\includegraphics[width=80mm]{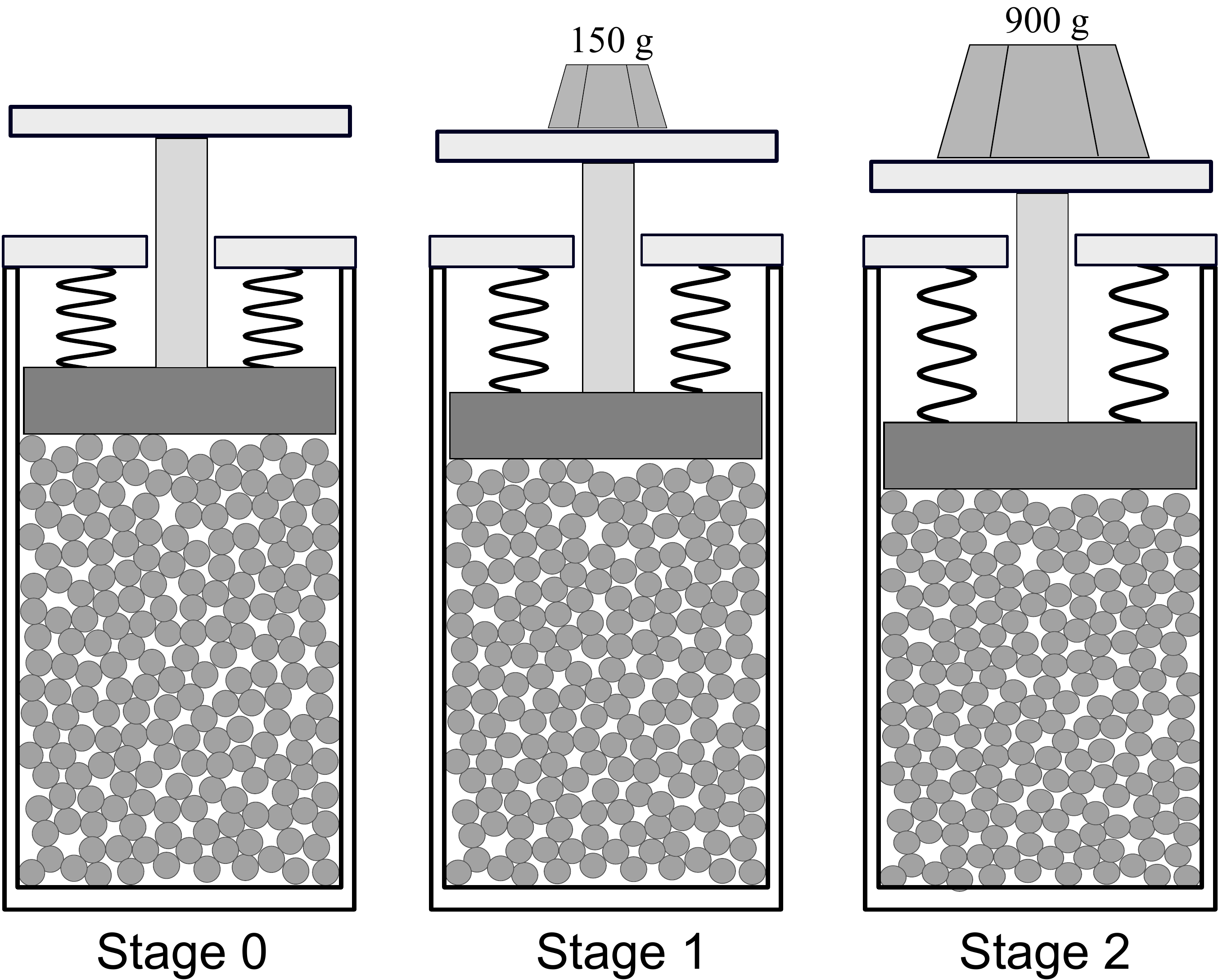}
\caption{Sketch of the compression cell and consecutive stages of compression.}
\label{ExperimetnSkech}
\end{figure}

The packing studied in this paper is made of relatively monodisperse spherical rubber balls of 
diameter $3.10\pm0.05~mm$ \footnote{Styrene-Butadine-Rubber (SBR); purchases from 
Mid-Atlntic Rubber Co., USA}. The maximum and minimum ball diameter 
measured in the packing are $3.19mm$ and $2.85mm$ respectively. 
The grains are made of commercial rubber with a shear modulus estimated at $850kPa$ .

A cylindrical PMMA container, with internal diameter of $44mm$ is used as a compression cell (see figure \ref{ExperimetnSkech}). 
The inner wall of the cylinder is lubricated with canola oil to reduce the friction. $2020$ rubber elastic balls are then poured 
into this chamber. The latter is then closed at the top by a piston of diameter slightly smaller than the container; it can 
therefore move freely 
without touching the PMMA cylinder and without allowing beads to leave the container. A horizontal platform is rigidly 
connected by a shaft to the piston so that a load can be applied to the sample by placing weights on it. A pair of 
springs connecting the piston 
to the container is used to make sure that, in the absence of a load, the piston does not fall down due to its own weight.

%The design of the cell, and in particular its dimensions, have been chosen in order to allow us to study the sample structure using X-Ray Computed Tomography (CT). 
The compression cell is attached to a motorised 
rotation stage located between a high-resolution microfocus X-ray source ($80 kV$ accelerating voltage and
$200\mu A$ beam current and a CCD detector of the size $67mm \times 67 mm$, $2048\times 2048$ pixels 
of size $33.6\mu m$ each \cite{Arthur:04a,Arthur:04b}. 
The compression cell is rotated by about $0.125$ degree increments around its vertical
axis and a radiographic projection of the packing is taken after each rotation with an
exposure time of $18$ s. A total of $2880$ projections are taken for a complete rotation of the
specimen. It takes approximately $15$ hours to scan the volume  for each case.  After the
completion of each scan, tomographic 
cone beam reconstructions are performed on these $2880$ projections using the a
Feldkamp algorithm \cite{Feldkamp84}. Tomograms of about $2000^3$ voxels are obtained at
$27$ microns resolutions. 

We have recorded and analysed the internal geometry of the packing under three different loading conditions. 

\begin{description}
\item[Stage 0:] Pre-compression stage with no extra loading; the piston is held in balance just above the 
top of the packing without making contact with it (see Fig. \ref{ExperimetnSkech}). Rubber balls are at rest, 
only bearing their own weight ($41.1g$). The combined weight of the piston and the 
platform ($600g$) is fully balanced by the pull of the springs so that at this point 
there is no external force being exerted onto the pile.

\item[Stage 1:] The platform is loaded with $150g$($1.47N$). By taking into account 
the opposite force applied by the springs, a net weight of $115g$ is applied to the 
packing, resulting in a compressive strain of $\epsilon=3.78\%$. We define   
the axial strains as the ratio of the displacement of piston to the 
initial height of the packing (engineering strains).

\item[Stage 2:] A total of $900g$($8.83N$) is placed on the platform resulting in a net $780g$ weight onto the packing. 
The total axial strain at that stage is $\epsilon=7.83\%$.
\end{description}

\subsection{3D Image analysis: segmentation and partitioning}

\begin{figure}
\centering
(a)
\includegraphics[width=60mm]{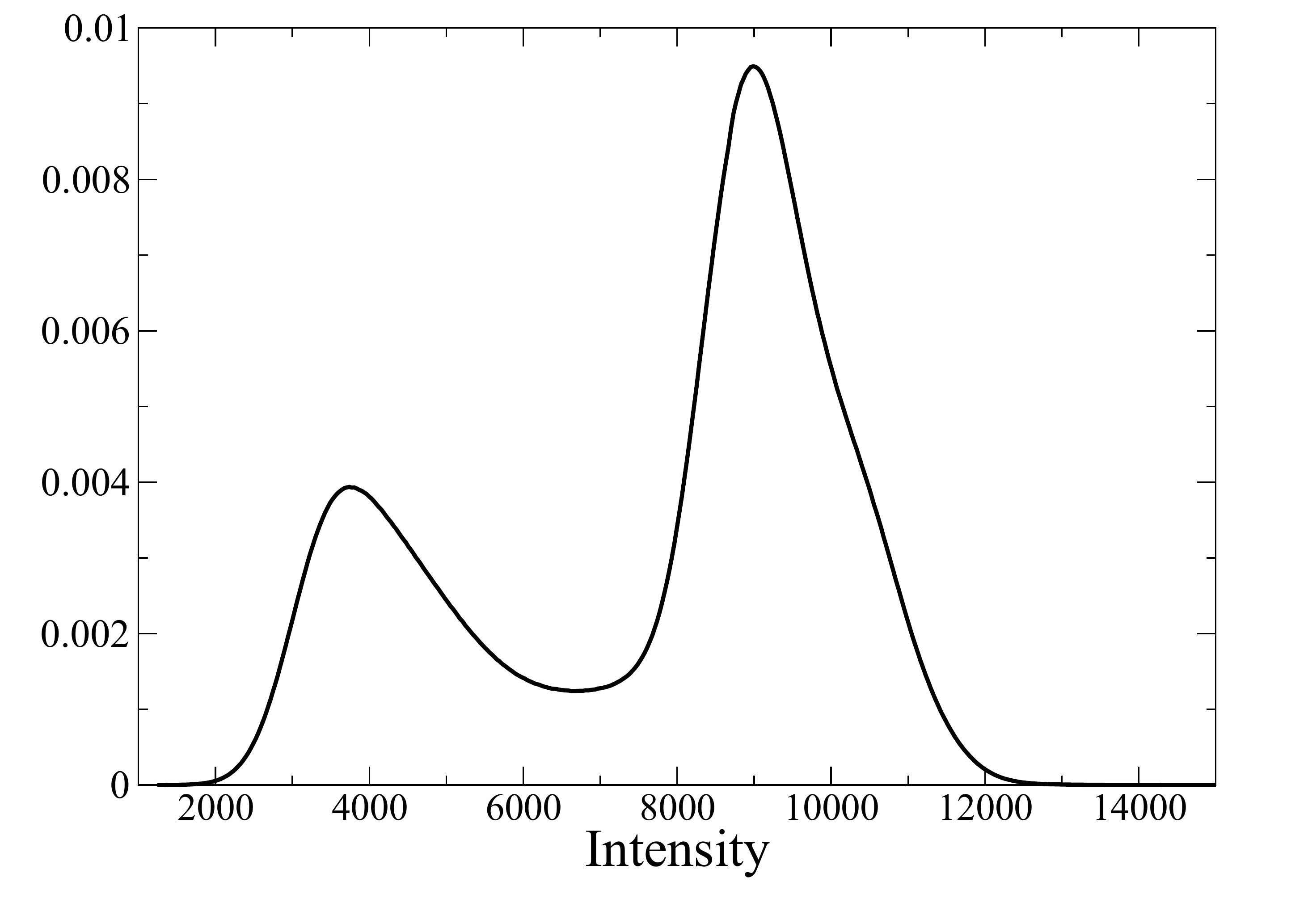}\\
\includegraphics[width=56mm]{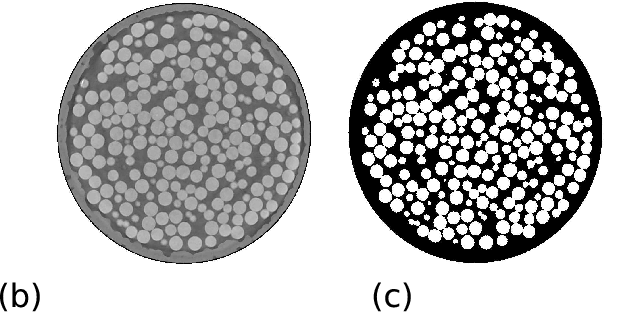}
\caption{(a) Normalized X-ray density histogram of the full image volume of the
rubber ball sample. (b) Gray scale X-ray density map of a slice of rubber ball
sample. (c) The same slices after phase separation.\label{histf}}
\end{figure}

The tomographic image consists of a cubic array of reconstructed linear X-ray
attenuation coefficient
values, each corresponding to a finite volume cube (voxel) of the sample. 
The first step in analysing this data is to
differentiate the attenuation map into distinct pore and grain phases. Ideally,
one would wish to 
have a bi-modal distribution giving unambiguous phase separation of the pore and
solid phase peaks.
This simple phase extraction is possible in our rubber ball pack. The intensity
histogram of the tomogram 
(Fig.~\ref{histf}(a)) shows two distinct peaks associated with the two
phases (beads and air). 
The peak centred around $9000$ is associated with the beads and the lower peak
around $4000$ is associated with 
the pore phase. 

The segmentation algorithm \cite{Adrian:04} uses both the attenuation value and its gradient to detect 
interfaces between the two phases (grain-void). In particular, a sharp gradient of attenuation is required 
to detect an interface. This method, after adjustment \cite{Saadatfar:06}, provides a very robust detection of 
grain boundaries, but is of course limited to void spaces larger than the voxel size. In particular, if the gap between two 
grains is very small (order of a voxel), the density gradient across the gap between 
these two grains will not appear as steep as a typical grain-void interface. This 
can result in the detection of false contact regions in the segmented datasets which consequently 
gives rise to the following undesired morphological effects: i) detection of false contacts between 
grains, and  ii) the surface area of real contacts appear larger than their actual values. As 
detailled later in this paper, this effect can be corrected in spherical grain piles by offsetting the apparent contact area of the 
grains. 

\begin{figure}
\includegraphics[width=84mm]{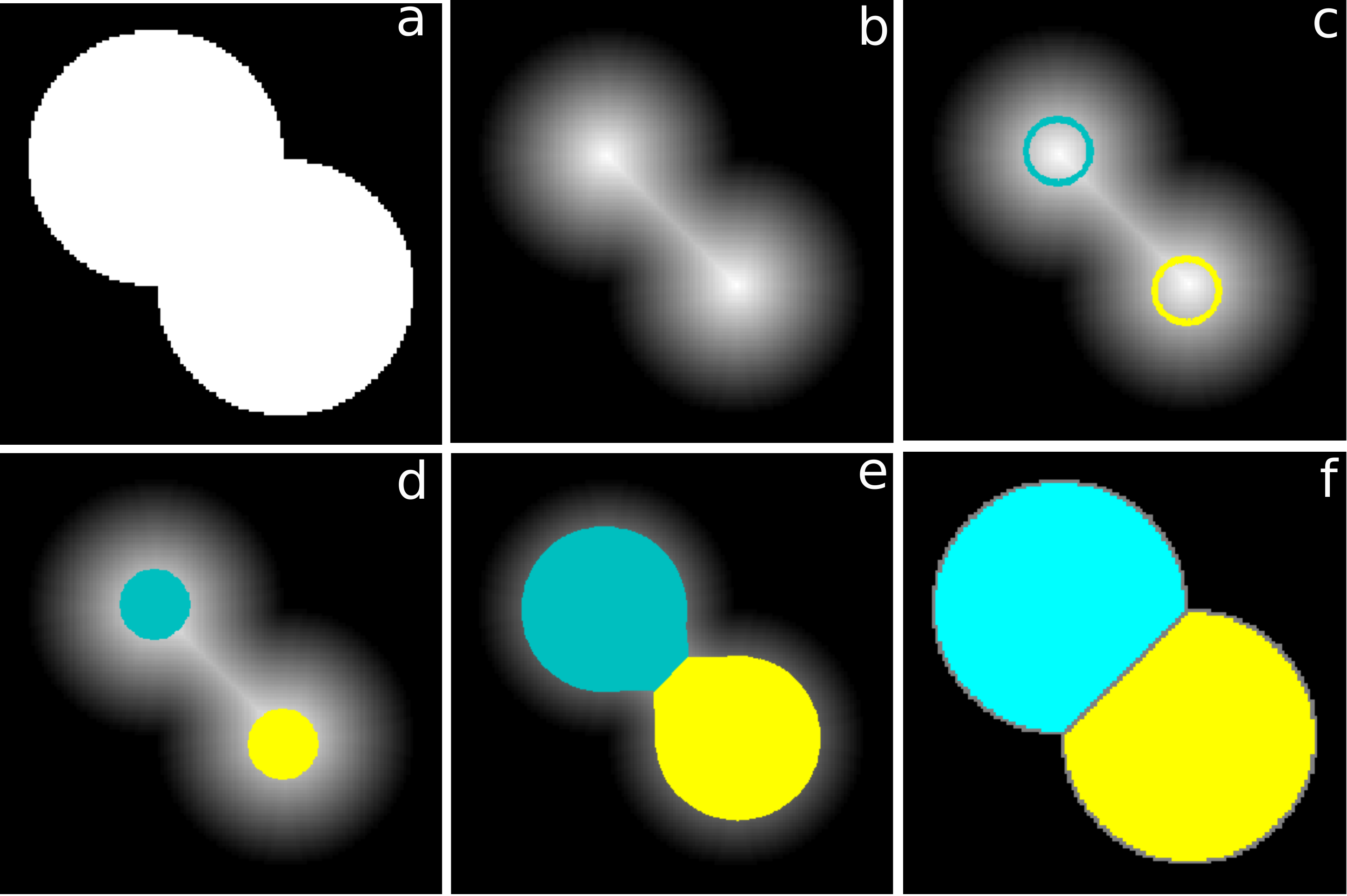}
\caption{Different stages of the watershed algorithm. (a) Two overlapping discs.
(b) EDT of the two 
overlapping discs. (c) Identifying the local maxima (seeds). (d) and (e) Regions
grow from the 
initial seeds by having pixels on their boundary added to them. (f) Discs
separated.\label{Watershed}}
\end{figure}

\label{PWA}

The next step is to reconstruct and label individual grains from the geometry of the solid phase. 
The basic assumption in the grain identification algorithm is that the boundaries 
between grains which are not isolated coincide with the watershed surfaces 
of the Euclidean distance map of grains (distance to the nearest grain boundary) 
\cite{Vincent:91}. The entire grain space can be thought of as the union of spheres centred
on every grain voxel. Each sphere radius is given by the Euclidean
distance value of the voxel at its centre. The next stage is 
identifying all the voxels that are not covered by any larger Euclidean
spheres. 

Each one of these voxels, which are at the maxima of the distance function in their 
local neighbourhood, then forms a seed that will grow into a single grain in the following
stage of the algorithm. Fig.~\ref{Watershed} illustrates a simple 2D case where 2 overlapping 
discs are separated. The seed regions essentially grow by 
having voxels on their boundary added to them. Voxels that lie on
the boundaries of the regions are processed in reverse Euclidean distance
order, i.e. voxels with high distance values are processed first. When a
voxel is processed, it is assigned to the region on whose boundary it
lies, or, if it lies on more than one region boundary, the region whose
boundary it first became part of. At the end of the algorithm, the grain 
space will be partitioned into grains whose boundaries lie on the
watershed surfaces of the Euclidean distance function \cite{saadatfarthesis}. 
The watershed based grain partitioning algorithm is computationally expensive. Hence it is
parallelised (using an implementation of the "time warp" discrete event
simulation protocol~\cite{Soille:03}), so it can be used to 
analyse very large datasets ($\sim 2000^3$).

\section{Measurements}

\begin{figure}
\includegraphics[width=70mm]{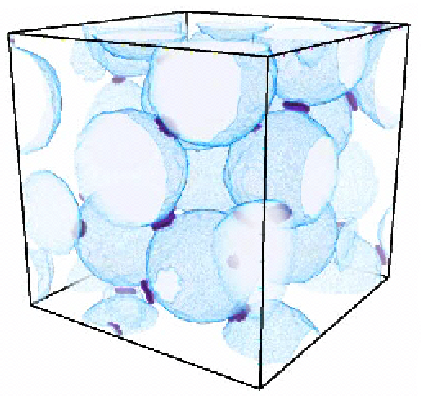}
\caption{A subvolume of the elastic ball compaction experiment. 
 The contact zone between touching grains (dark patches)  
 provides a measurement of contact force.\label{RBContactForce}}
\end{figure}

The 3D watershed algorithm provides us with large amounts of information about the pile structure.
Each grain has been labelled, its location and shape are known, its direct neighbours have been identified, 
and the geometry of the contact they share can been extracted as well.
This data can now be processed to extract physical and mechanical properties. In this
section, we present a few methods implemented in the
context of granular systems. These can be separated into structural characteristics and 
mechanical measurements. The latter involves kinematic and force measurements 
that rely on prior knowledge of a contact law for individual grains.

\subsection{Structural quantities}

\subsubsection{Packing fraction}

One of the most readily measurable quantities from 3D images is the packing
fraction. Packing fraction or apparent density is 
defined as the ratio of the volume of balls to the total volume. 
Table~\ref{Tab1} summarises the packing fractions measured directly from the
digital images. The packing fractions we measure on the initial pile is $57\%$, 
which is in the lower end of the spectrum for monodisperse beads. Such a density 
is achievable in our system due to the large frictional coefficient of rubber-rubber 
contacts. As the loading is increased, the volume fraction increases up to $62\%$.

\begin{table}[b]
\centering
\begin{tabular}{|c|c|c|c|}
\hline
Compression stage              & $0$ & $1$     & $2$           \\  \hline \hline
Loading [gr]                   & $0$ & $115$   & $780$         \\ \hline
Strain  [\%]                   & $0$ & $3.78$  & $7.83$        \\ \hline
Packing fraction               & $0.57$ & $0.59$ & $0.62$   \\ \hline
Avg. coord. no.                &  \multirow{2}{*}{$6.28\pm1.64 $}  & \multirow{2}{*}{$6.63\pm1.58$} & \multirow{2}{*}{$7.13\pm1.49$}    \\
raw & & &\\ \hline
Avg. coord. no.                & \multirow{2}{*}{$4.99\pm1.80$} & \multirow{2}{*}{$5.32\pm1.66$}&\multirow{2}{*}{$5.92\pm1.56$} \\ 
filtered & & & \\ \hline
\end{tabular}
\begin{center}
\caption{Details of the compression progression.}
\label{Tab1}
\end{center}
\end{table}

\subsubsection{Radial Distribution Function} 

\begin{figure}
\includegraphics[width=70mm]{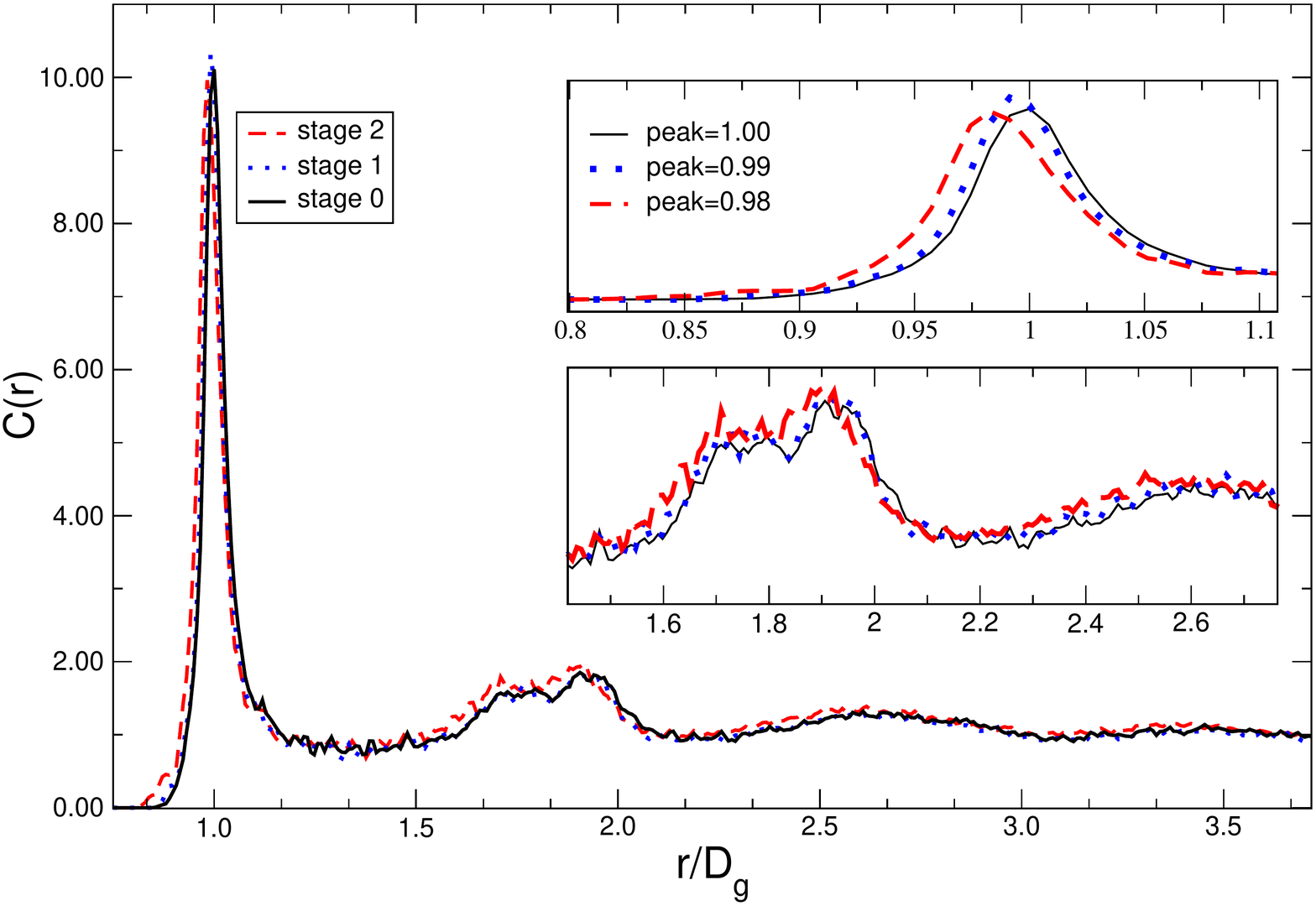}
\caption{Autocorrelation function of the packing at different 
stages of compression normalized by the volume of a single grain 
and the packing fraction. Note the shift in the peaks as the pressure increases.
\label{2pcf}}
\end{figure}

From the shapes of the individual grains calculated by the Watershed algorithm, 
the coordinates of each grain's centre of mass can be accurately measured. This 
data is in particular required to characterize the microstructure of granular
systems and to calculate the autocorrelation function of the grain centres.

The radial distribution function (RDF) is a measure of the degree of separation of 
grain centres and their density at a given distance. It is calculated by counting 
the number of grains $N$ that are separated from a given sphere by a distance in the 
interval $[r, r+dr]$, where $r$ is the distance between grain centres. 
For large $r$ the number density of the grain centres found in the interval $[r, r+dr]$ 
approaches the average density of grains in the packing, $\frac{3}{4\pi {R}^3}\rho$, where 
$\rho$ is the packing fraction and $R$ is average grain radius. In our calculations, 
we normalize RDF by the average density. Figure \ref{2pcf} shows the RDF in our 
packings throughout compression progression. The RDF of the system before insertion of any external
force (stage 0) shows a prominent peak at $r\approx D_g$ where $D_g$ is the average grain
diameter ($D_g=2\times R\approx3.10 mm$). The second peak appears at $r\approx1.95D_g$ and a sub-peak 
approximately at $r=\sqrt{3}D_g$. As the vertical load increases (stage 1 and
stage 2), the peak of the RDF widens and shifts to the left due to the compression of touching
spheres, hence shortening the centre-centre distance between them (see Fig. \ref{2pcf}(left)). 

\begin{figure}
\centering
(a)
\includegraphics[width=70mm]{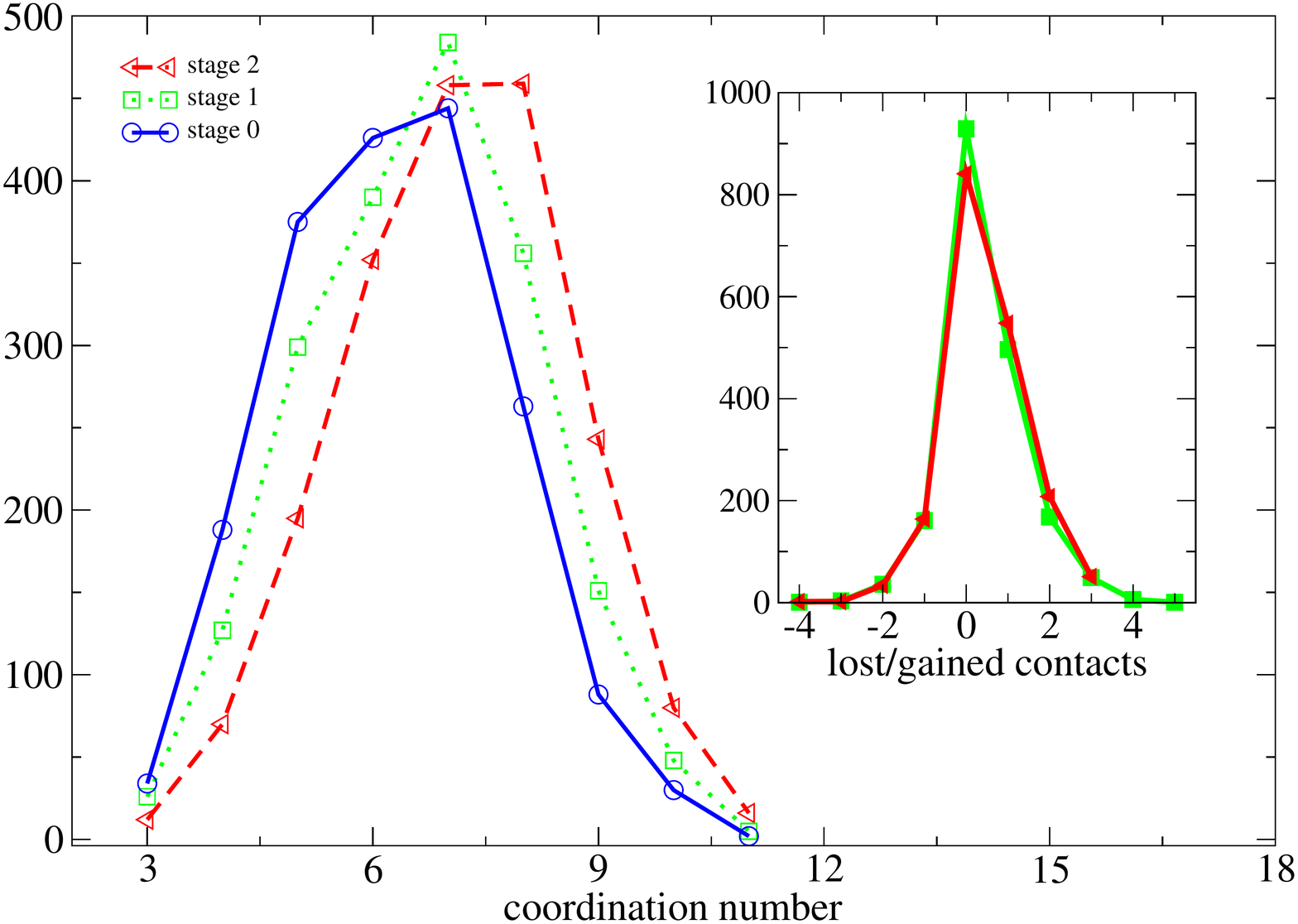}\\
(b)
\includegraphics[width=70mm]{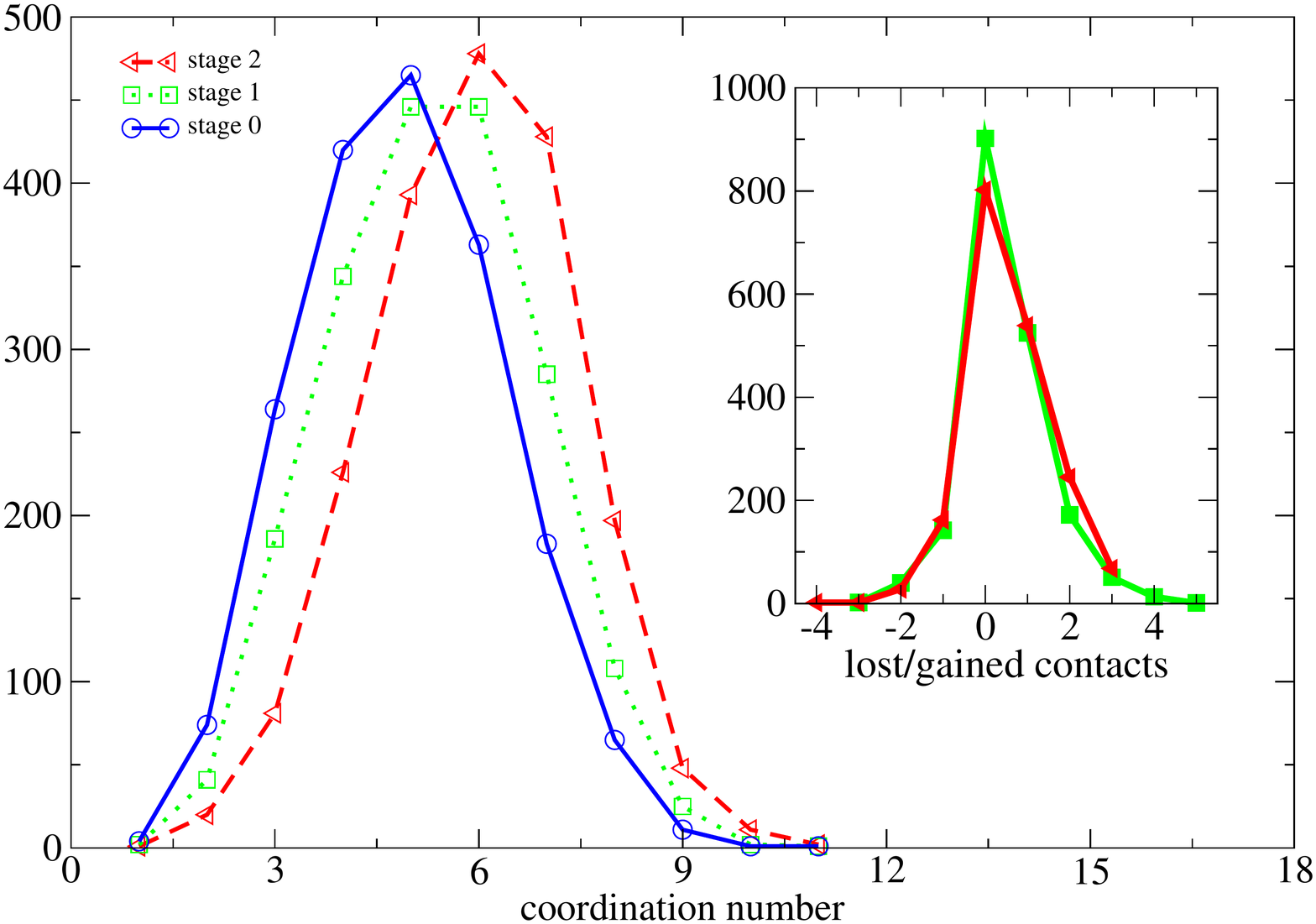}
\caption{Histogram of coordination number as a function of loading for
different threshold values: (a) No threshold is applied (using the raw data after watershed algorithm). (b) After a threshold value 
of $0.35~mm^2$ is subtracted from all contact surface areas. As a result 
the average coordination number is reduced in all three stages.
\label{Z}}
\end{figure}

The magnitude of the shift is about $1\%$ and $2\%$ for stages $1$ and $2$. 
This is to be compared with the macroscopic axial strain which is of the other 
of 4\% and 8\% without any deformation along the order directions. This suggests 
that the internal compression is more isotropic than the loading.

\subsubsection{Mean coordination}
\label{meancoordnumber}

The grain partitioning also provides us with the list of contact areas between grains, from which it is possible to analyze the contact network of grains.  
Using first the raw data from the watershed algorithm, we have measured the coordination (number of contacts) of each grain and plotted its distribution for 
each stage of compression (see Figure \ref{Z}(a)). This plot shows a significant
increase of grain coordination as the loading increases. It is notable that as pressure increases,
the distributions move to higher values and get slightly narrower (see Table \ref{Tab1}). This increase 
in the coordination number is achieved by re-organisation of grains in the packing when they are compressed 
and also grain compression/deformation process which reduces the overall grain-grain distance. 
A relatively large number of grains lose their contacts with some of their immediate neighbours 
while almost the same amount gain new contacts as the compression progresses. We have shown the histogram 
of lost/gained contacts in the inset of figure \ref{Z}. Negative values on the horizontal axis represent lost 
contacts while positive values show the number of newly gained contacts. Nearly $50\%$ of grains 
retain their coordination number during the compression without any changes.

\begin{figure}
\includegraphics[width=80mm]{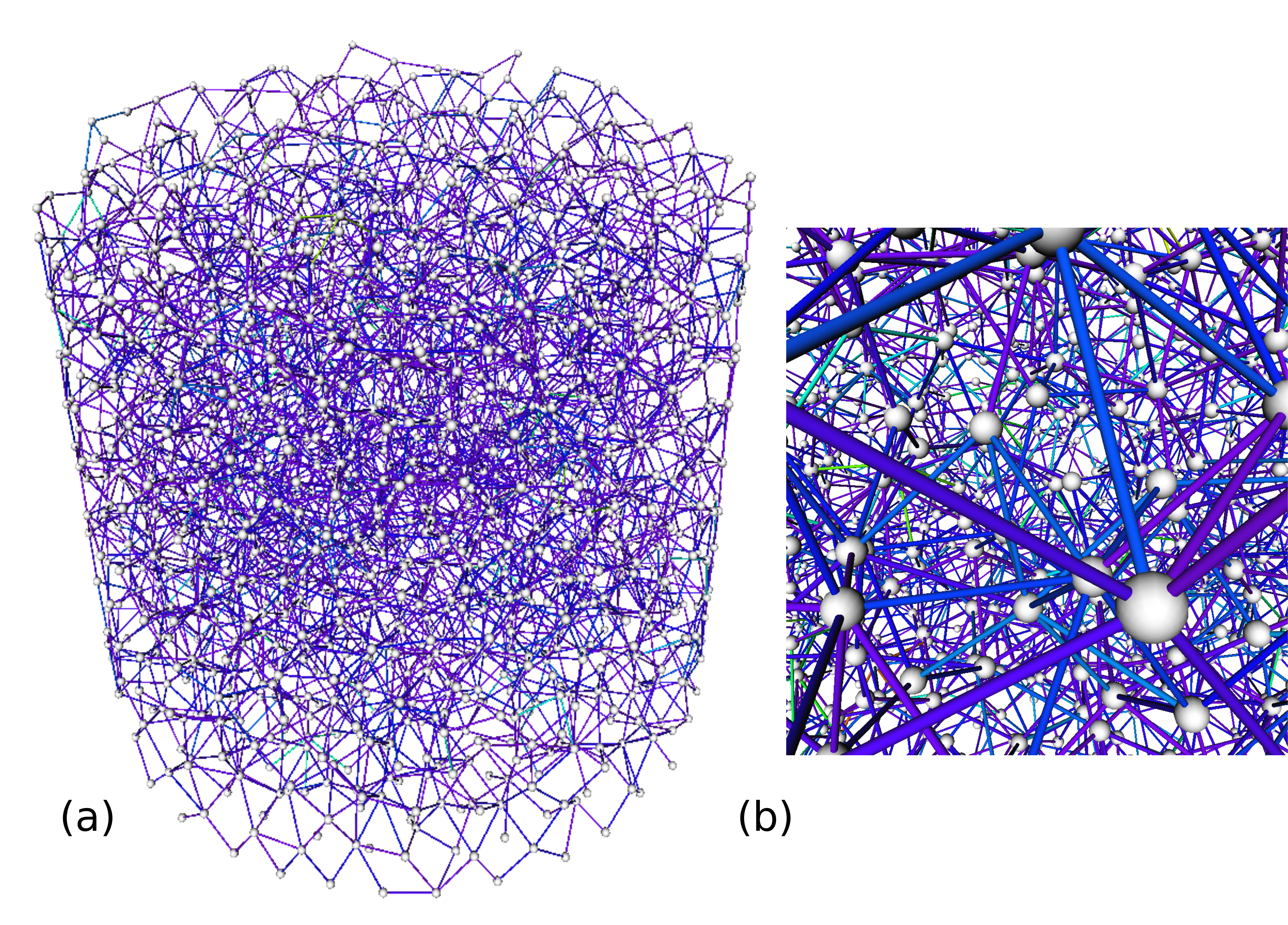}
\centering
\caption{(a) Network representation the full packing. (b) An internal view of
the packing illustrating the connecting grains
\label{Network}}
\end{figure}

As discussed in section \ref{PWA}, the inherent finite resolution of the CT leads to systematic bias in 
fine details such as grain contacts; if the gap between two grains is in the order of, or less than, the 
voxel size, grains will appear to be in contact in the segmented data. To correct for this intrinsic resolution 
limitation, we offset all the contact areas by a small amount, which essentially corresponds to the apparent 
contact area of two touching perfectly spherical grains (see section \ref{forcedist} for detailed discussion). 
All contacts whose apparent surface area is below this threshold, are therefore discarded. Figure \ref{Z}(b) 
shows a re-plot of figure \ref{Z}(a) after applying the offset. The isostatic limit for mechanically stable 
packing of spheres in 3D suggests a connectivity of $4$ for frictional and $6$ for frictionless systems. 
In our measurements, the average coordination number is $\simeq5$ for stage $0$ (see Fig.\ref{Z}(b)) which 
is in agreement with previous measurements \cite{bernal:60,Aste:05}.
 
From the knowledge of the whole contact network, a large number of other morphological and topological quantities can be calculated, 
such as spatial correlations in contact orientations, or contact anisotropy. Figure \ref{Network} shows a 
reconstruction of the contact network through which forces propagate. 
The statistical properties of such networks will be developed in 
further studies. In what follows, we will focus on the determination of 
mechanical forces within the pile, at the micro scale, as well as 
mesoscopic scale.

\subsection{Kinematics and dynamics}
\subsubsection{Displacement fields}

\begin{figure}
\centering
\includegraphics[width=80mm]{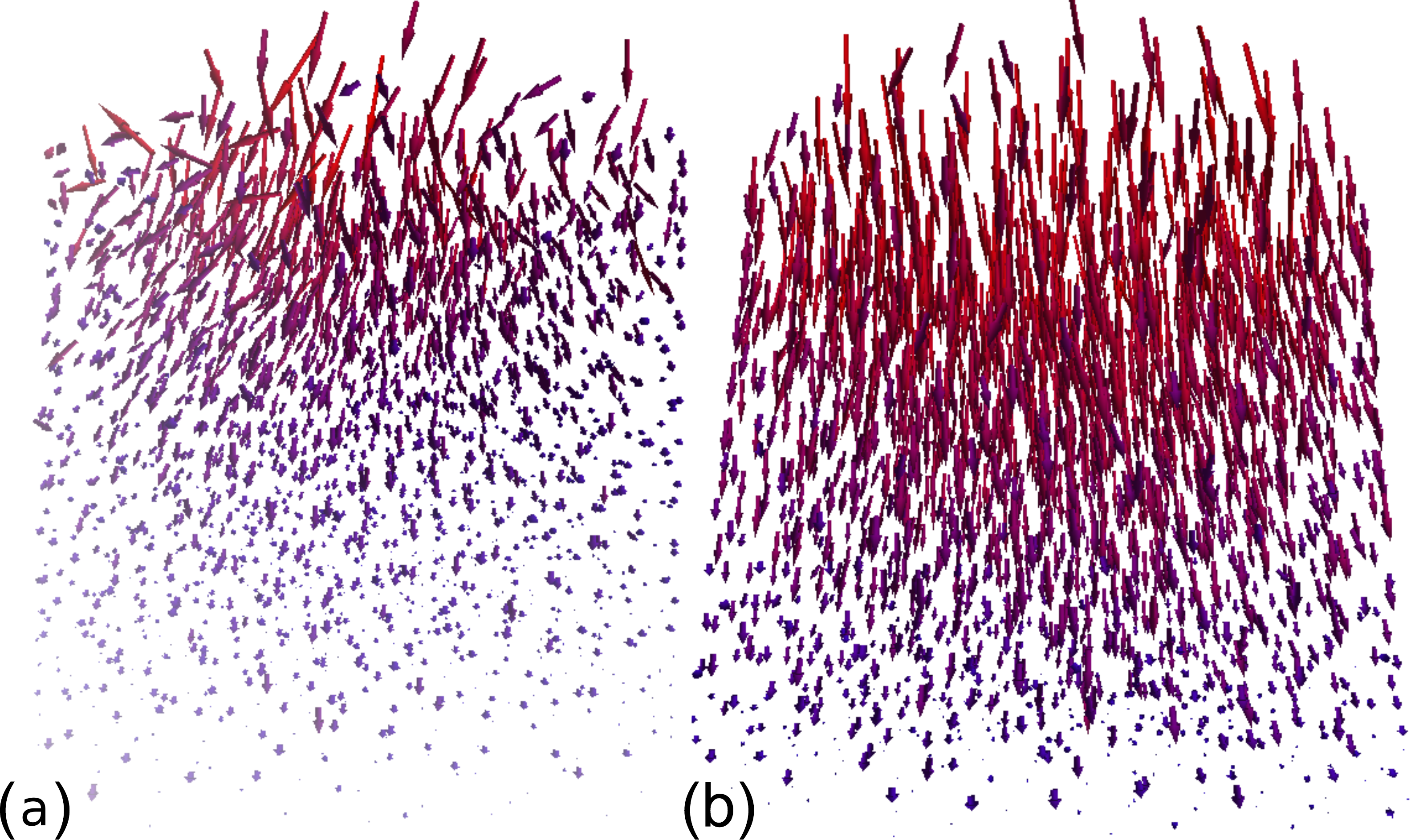}
\caption{Displacement field presented as vector fields (velocity fields) for stage 1 (a) and 2 (b) of the experiment.\label{DisplacementField}}
\end{figure}
 
We calculate next the coordinates of the centroid of the grains for all three
scans and then track each individual grain 
throughout the experiment. As an illustration, we render the 3D vector field
derived from tracking grains during the compression for 
both stages of compression, see figures \ref{DisplacementField}(a-b).

\begin{figure}
\centering
\includegraphics[width=75mm]{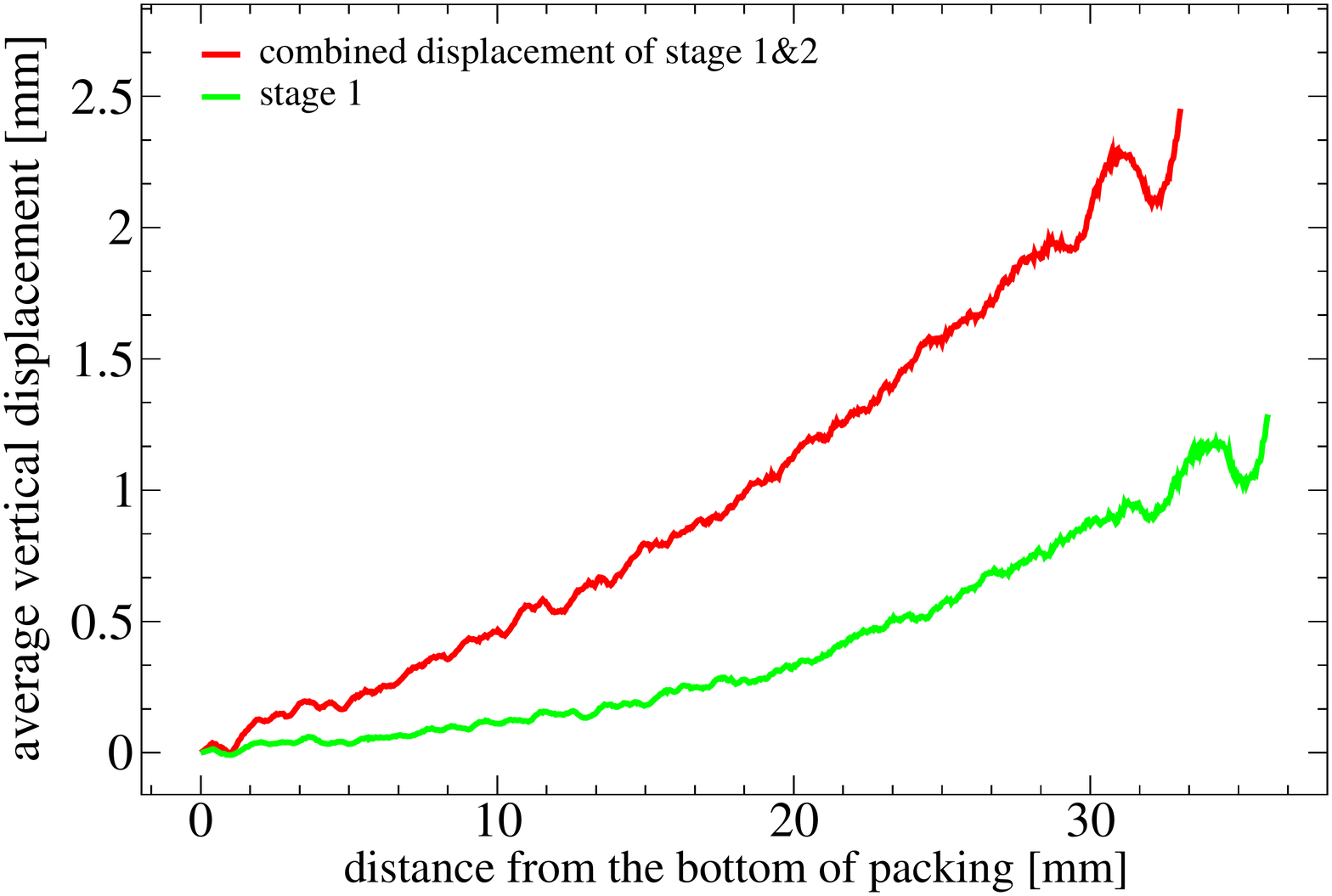}
(a)
\includegraphics[width=75mm]{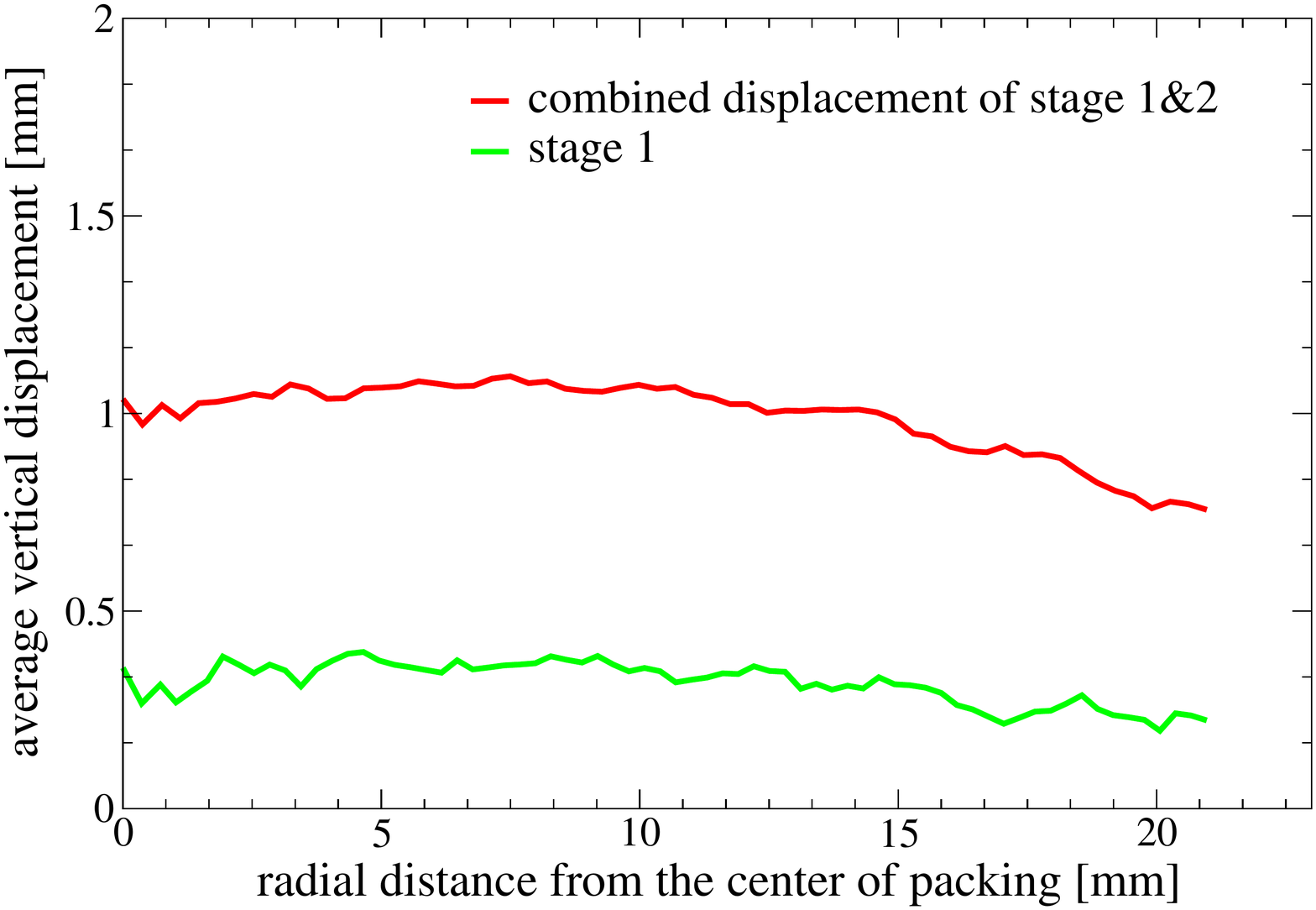}
(b)
\includegraphics[width=75mm]{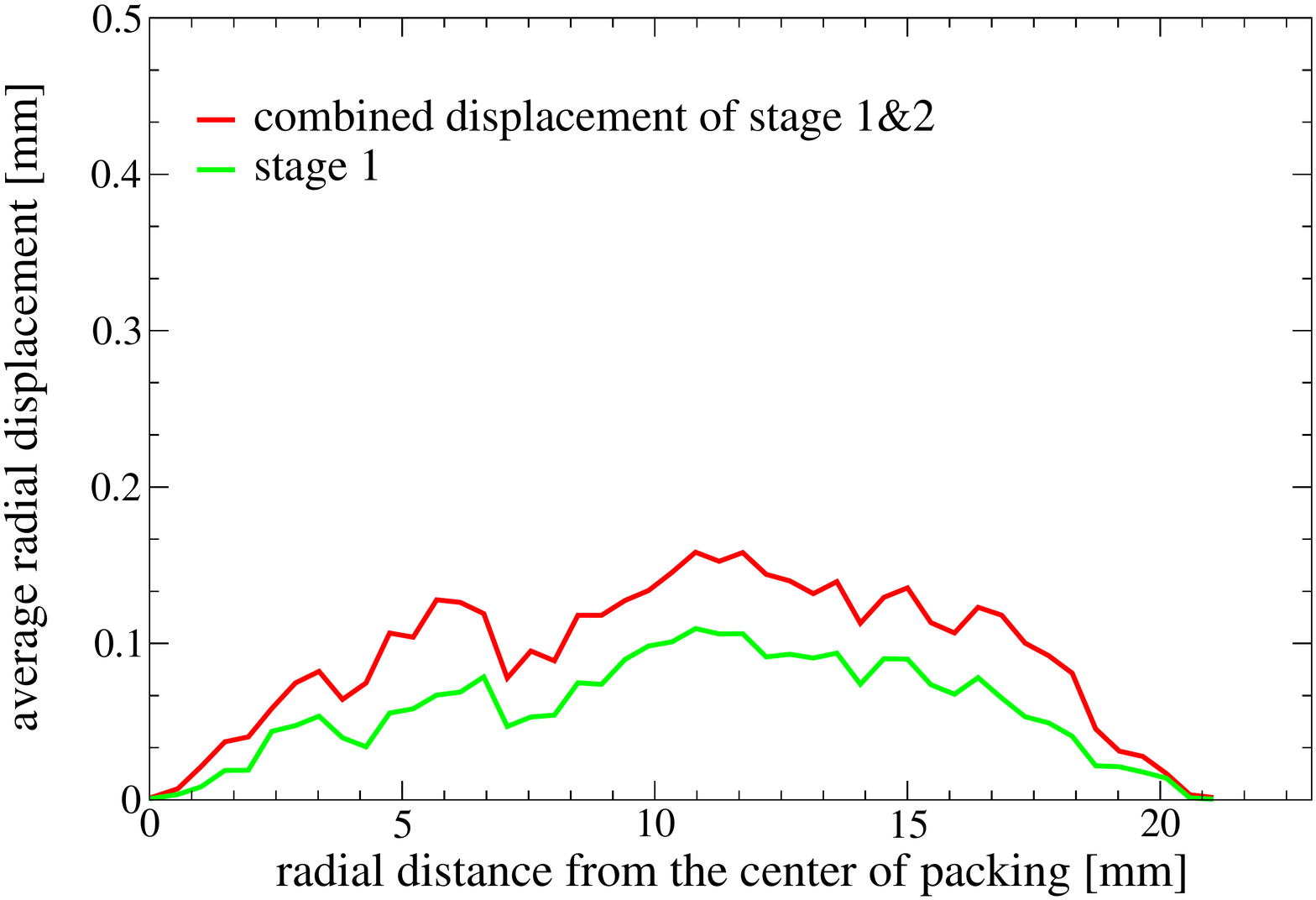}
(c)
\caption{Displacement field along the loading axis. (a) The overall vertical displacement field is 
represented very well by an exponential equation. For visual convenience every 30 data 
points is shown. (b) Longitudinal displacement for the 1st stage and the overall 
displacement. Here we choose a bin size of $0.4mm$ for averaging. (c) Average radial 
displacement of grains measured from the distance between the centre of packing its outer 
boundary.}
\label{Displacement}
\end{figure}

To gain a qualitative insight into the displacement field, we measure the 
average displacement of grain centres in both longitudinal and radial directions. Figure
\ref{Displacement}(a) demonstrates such displacement along the loading axis. 
As expected, gradient of displacement from top to bottom shows that the
system is under vertical compression. The decrease of the slope with depth also 
indicates that part of this stress is screened by the time it reaches the bottom. 
A typical explanation of this is the Janssen effect \cite{Janssen:1895}, suggesting 
that friction on the lateral walls might be significant. To confirm this, we studied 
the vertical displacement profile in the radial direction (figure \ref{Displacement}(b)) which 
confirms that grain movements closer to the outer region of cylinder are smaller than that 
of the central region. Therefore a shear component in the $(z,r)$ direction is expected. 
Another feature of the displacement field is the weak but noticeable radial displacement 
of the grain towards the boundary (see figure \ref{Displacement}(c)). This radial displacement 
can be clearly seen in figure \ref{DisplacementField}(a).

\subsubsection{Contact mechanics}

In addition to the direct measurement of individual grain displacements,
a number of other mechanically relevant features can be extracted from the
tomogram. Upon compression, we observed that the distance between grains
varies, as well as shape and contact areas between grains. These 
quantities can {\it a priori} be used to quantify the force 
transmitted between the grains. In this section, we describe the implementation of a contact
model that is suitable for stress calculations from 3D images.

\begin{figure}
\centering
\includegraphics[width=80mm]{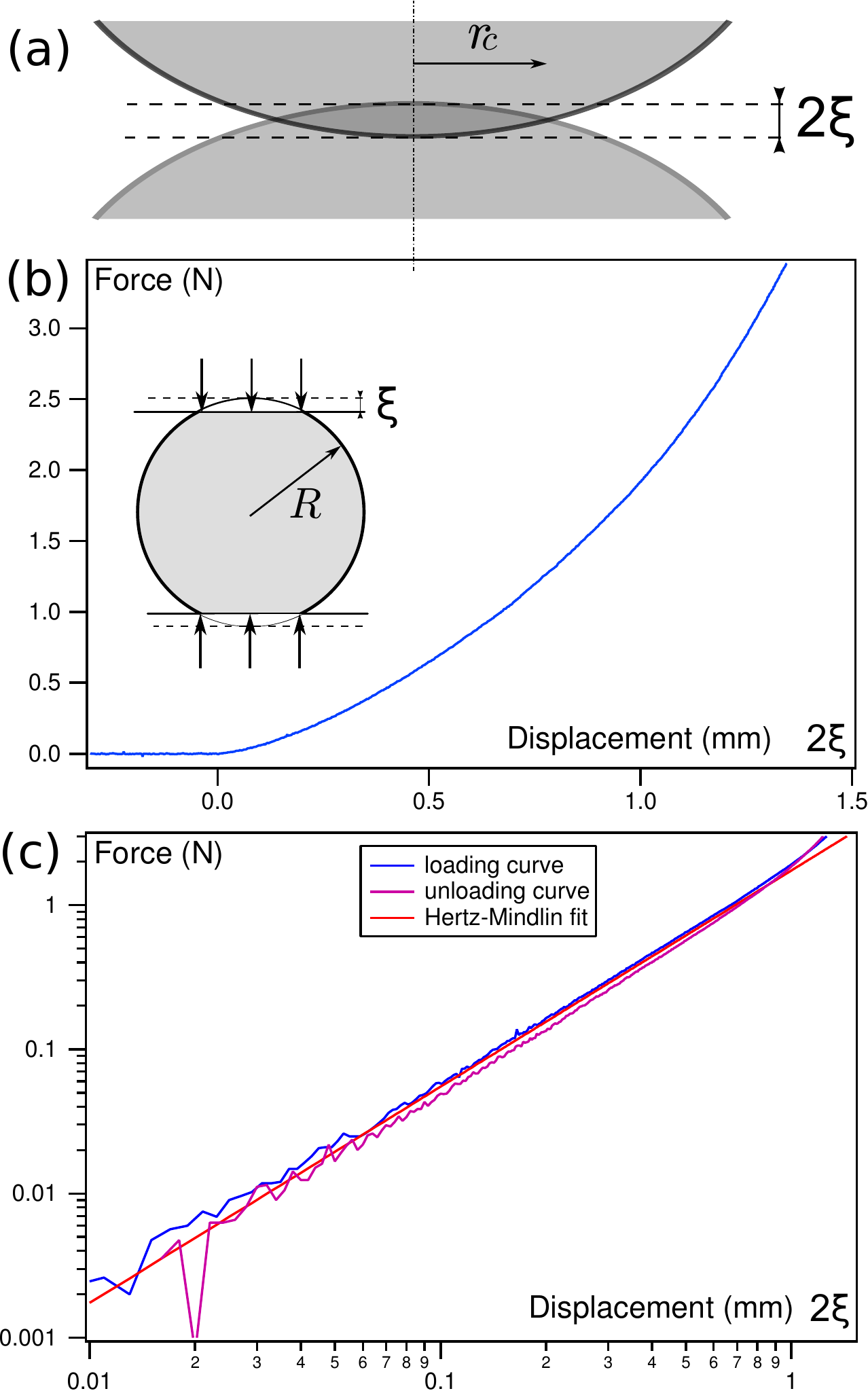}
\caption{Response of a single rubber ball to loading.\label{singlebead}}
\end{figure}

Contact mechanics between solid objects have been studied for over a century. 
The seminal works of Hertz \cite{Hertz:1882}  and Mindlin \cite{Mindlin:49}
about the contact mechanics of solid bodies have provided analytical solutions for idealised cases, 
summarised in \cite{Johnson:87}. The contact force between two elastic spheres can be calculated from the
nonlinear Hertz-Mindlin model. The expression of the normal force between
two contacting elastic spherical grains $m$ and $n$ with uncompressed 
radii $R_m$ and $R_n$, made of the same material, is given by:

\begin{equation}
f_{mn}=\frac{2}{3} \frac{4G}{1-\nu} R^{1/2}\xi_{mn}^{3/2},
\label{NormalForce}
\end{equation}

where $R$ is the geometric mean of $R_m$ and $R_n$, $R=2R_mR_n / (R_m+R_n)$,
$\xi_{mn}$ is the normal overlap, or penetration length, as depicted in Fig. \ref{singlebead}a, 
$G$ is the shear modulus and $\nu$ the Poisson ratio of the grains, 
equal to $0.5$ for rubber \cite{Zhang:05a}. 

\begin{figure}
\centering
\includegraphics[width=80mm]{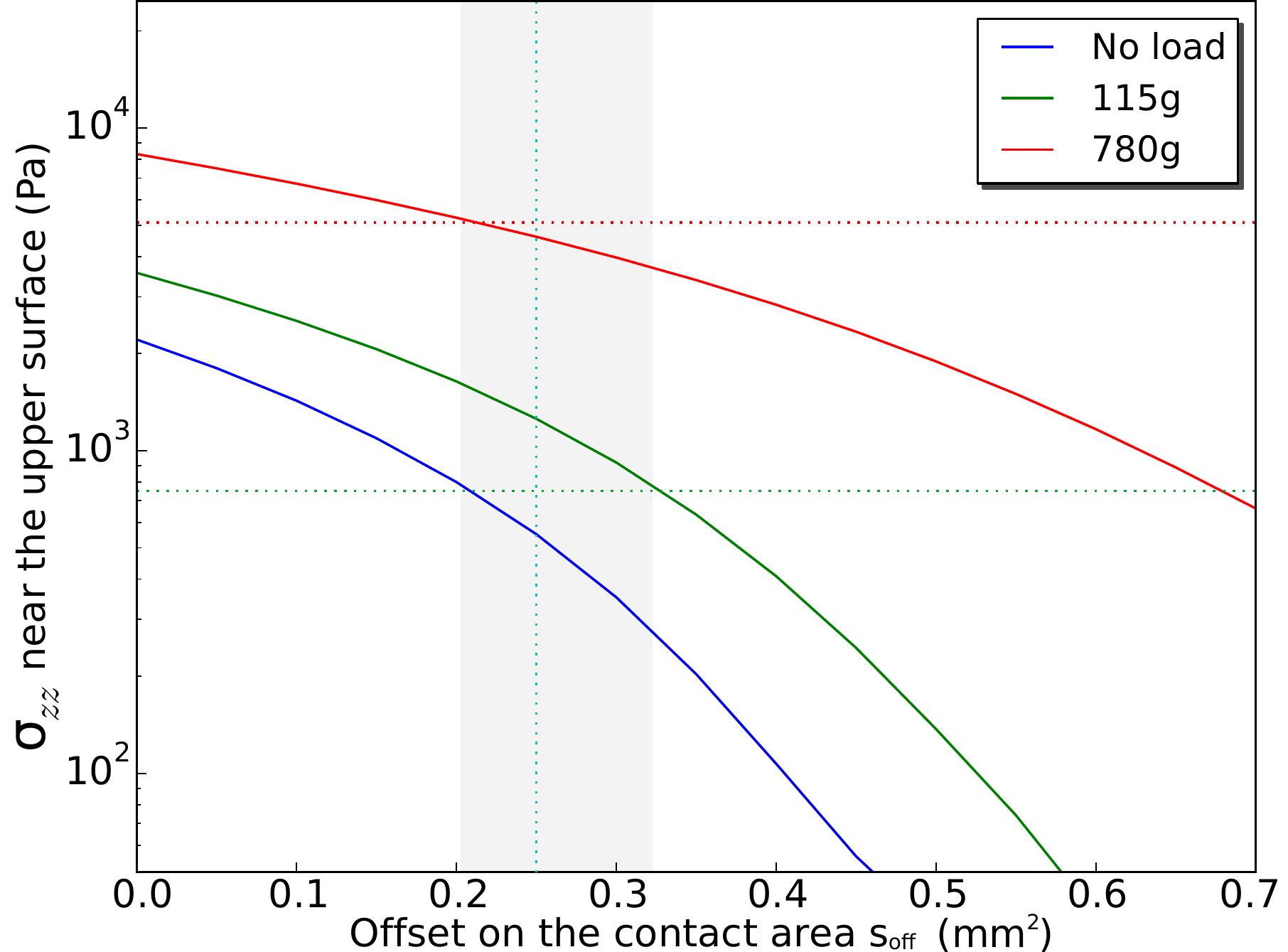}
\caption{Solid lines: the mean vertical stress as a function of 
contact area offset. Dotted lines: the applied load, 740 Pa for stage 1, and 5000 Pa
for stage 2.
\label{StressOffset}}
\end{figure}

We have tested the mechanical response of our grains by compressing a few individual grains between two steel plates and measuring 
the force required as a function of the gap between the plates (Fig. \ref{singlebead}). 
The response of the grains is consistent with the Hertz model, and we have extracted from these graphs the 
value of the grains shear modulus, $850kPa$, which is a reasonable value for a commercial rubber.

The Hertz-Mindlin theory therefore provides us with a suitable model to calculate the contact force from the grain geometry, as 
long as we have a good estimate of the normal overlap $\xi$ for each contact. Two options can be considered at this stage. i) The normal overlap can 
be estimated from the distance between grain centres. If the grains are located at 
positions $\underline{r}_m$ and $\underline{r}_n$, assuming grains remain spherical, $\xi_{mn}=\frac{1}{2}[R_m+R_n-|\mathbf{\underline{r}_m-\underline{r}_n}|]$.
The autocorrelation function shows that we expect this quantity to be of the order of a few percent of the grain diameter. 
Although the grain location is determined accurately, calculation of $\xi$ can be fairly inaccurate due to the anisotropy of the imposed strain causing the 
grains to deviate significantly from a spherical shape into more ellipsoidal geometries. The 
centre-centre distance is therefore not a good approach to estimate the contact geometry.  ii) Our image processing 
protocols deliver a sensitive measurement of the contact area (Fig. \ref{RBContactForce}), 
which provides us with a more reliable way to calculate the forces independently of the centre-centre distances. 
The contact area $(s_{mn})$, between grains labelled by $m$ and $n$, is related with the normal overlap $\xi_{mn}$ by: $\xi_{mn}=\frac{s_{mn}}{2 \pi R}$. 
This purely local measurement of the force can be applied to any grain geometry, as 
long as the local curvature of the grain near the contact point can be estimated as well.

\subsubsection{Force distribution and stress field}
\label{forcedist}

As discussed above, the finite resolution of the CT causes the contact areas to
be systematically larger than their real values so this measurement needs to be 
carefully calibrated. There is {\it de facto} a small distance $\delta$, of the order of the voxel size, such that if 
two surfaces are separated by less than $\delta$, they will appear in contact after segmentation. This results in a 
systematic enlargement of the contact area, and even detection of false contacts if the separation between grains is 
less than $\delta$. In the case of spherical elastic grains, the contact area after segmentation would correspond to 
the geometrical overlap of two grains with each radius being increased by $\delta/2$. Since the contact area is linear 
with the contact overlap $\xi$, the effect that $\delta$ has on the surface area is equivalent to a systematic offset 
$s_{off}=\pi R \delta$. However, $\delta$ is a priori unknown and has to be determined by calibrating our measurements.

\begin{figure}
\centering
\includegraphics[width=80mm]{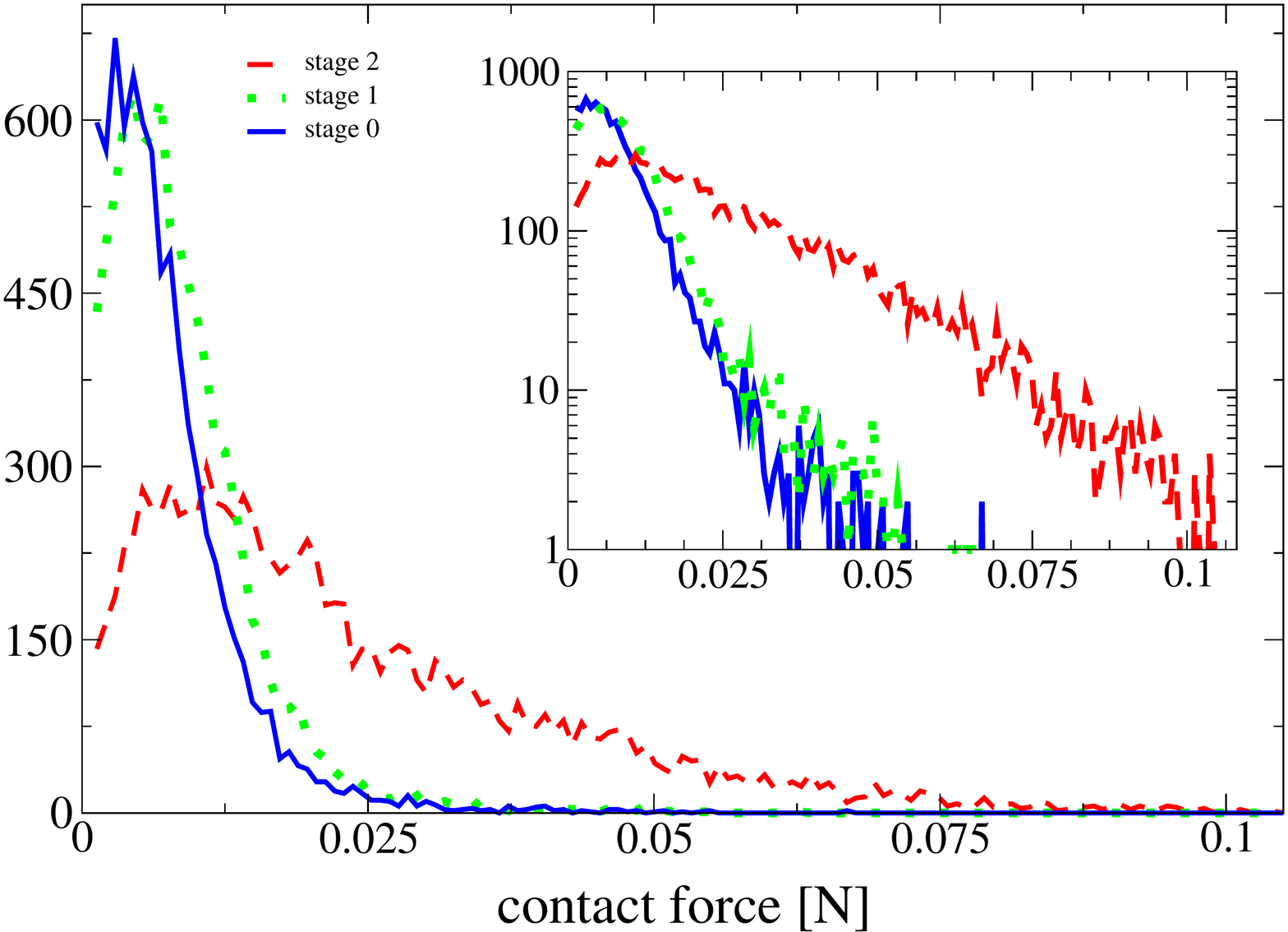}
\caption{Histograms of normal forces in the sample for all three loading
stages.\label{forceHisto}}
\end{figure}

The mean stress in the vertical direction at the vicinity of the upper interface can be estimated in two independent ways for 
each loading step; either from the knowledge of the loading mass and geometry of the setup, or by using the  
contact forces to calculate the stress tensor in the bulk of the pile. We consider a subvolume $\delta v$. 
%It contains  a set of contacts between pairs of  grains indexed by $m$ and $n$. 
Each contact contained in this volume bears a force denoted $\mathbf{\underline{f}_{mn}}$, 
measured from the contact area $s_{mn}$ between grains indexed by $m$ and $n$. The 
centre to centre vector is noted $\mathbf{\underline{l}_{m,n}}$. The components of the stress 
tensor, indexed by $i$ and $j$, are then obtained from the following expression:

\begin{equation}
\sigma_{ij} = \frac{1}{\delta v} \sum_{m,n} \;f^i_{mn} l^j_{mn}
\end{equation} 

\noindent where the sum is over all contacting pairs of grains in the volume $\delta v$.

In order to measure the offset $\delta$ required to compensate for the segmentation error, 
we have calculated, using the upper third of the sample, the mean vertical stress $\sigma_{zz}$ 
we would obtain for a range of values of the surface offset $s_{off}=\pi R \delta$ (Fig. \ref{StressOffset}). 
A suitable choice should provide the value of the normal stress consistent with the loading applied on the sample (\textit{i.e.} 
where the solid line intersects the dotted line on Fig. \ref{StressOffset}). Based on this graph, we select 
$s_{off} \approx$ 0.25 mm$^2$, that nearly sets the normal stress to its expected value for the high load value, 
where the calibration is the most reliable due to the large values of the force and increased number of contacts. 
This value corresponds, as expected, to the size of a single voxel, confirming the consistency of the method. For 
stage 0 and 1, however, based on this single threshold calibration, the overall mean stress 
appears to have been overestimated.

\begin{figure}
\centering
\includegraphics[width=70mm]{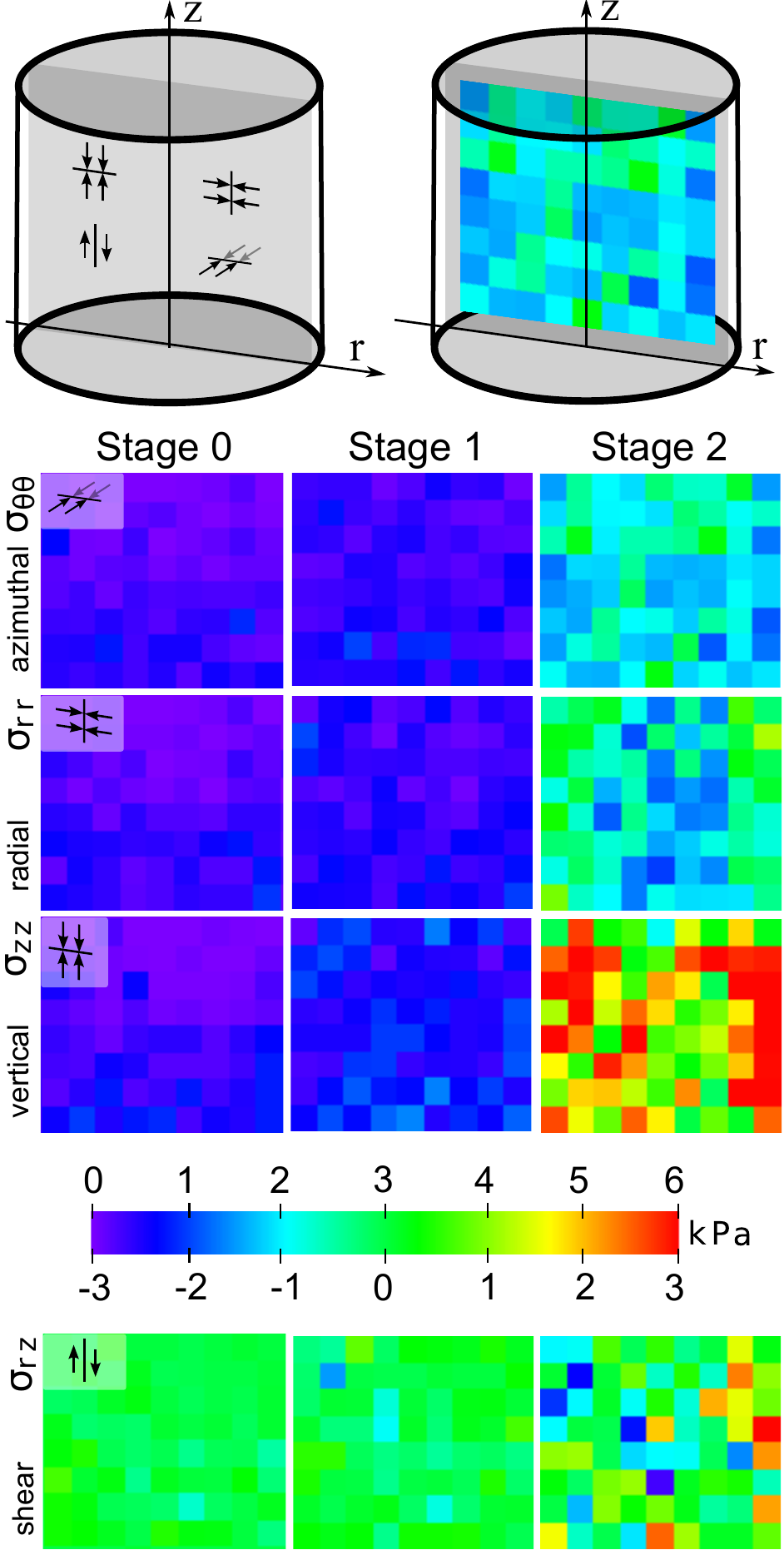}
\caption{Stress fields in a plane section of the sample for various loads. From 
top to bottom, i) cartoons showing the location of the section in the sample, 
along a diameter, and various stress components reported, ii) the
stress map in the azimuthal direction $\sigma_{\theta \theta}$, in the radial
direction $\sigma_{rr}$, in the vertical direction $\sigma_{zz}$, and shear
stress $\sigma_{rz}$.  \label{stressfig}}
\end{figure}

% \begin{figure}[t]
% \centering
%
%{\includegraphics[width=.42\textwidth]{./Plots/2pcf_RB-TwoInsents-ReviewPaper}}  \hfil
% {\includegraphics[width=.4\textwidth]{./Plots/RubBalls-Z-percent}}\\
% \caption{(left) The autocorrelation function of elastic balls system. Note the
%shift in the peaks as the pressure increases. 
% (right)  Grains with connectivity  two and three belong to the 
% boundary of system. }
% \label{RB-2pcf-Z}
% \end{figure}

Once this tool is calibrated, it can be used to probe a number of statistical quantities in the pile.
Figure \ref{forceHisto} shows the distribution of normal forces in the pile for the three different loads studied here.
The 3D bulk measurement of the contact normal forces exhibit a number of features that are characteristic 
of granular systems. In particular, a salient feature of force measurement presented in this study  
is that the distributions are primarily exponential for large forces. This is in agreement with earlier measurements in 
2D piles \cite{Coppersmith:97,Liu:95} or at a 3D interface \cite{Bruji:05} and also the 
numerical studies of dynamics of granulated systems \cite{Radjai:96}.  
Another feature often reported in granular dynamics studies \cite{OHern:01,Majmudar:05} 
is the presence of a peak or a plateau at low forces. However, this region of the distribution is also where the 
forces are the most sensitive to image resolution, segmentation and thresholding. We believe a much higher spatial resolution is 
required before conclusions can be made about the precise shape of the force distribution in the region 
of low forces, in particular since the mean stress could not be adjusted.

The stress field in the pile is represented by a series of color maps in Fig. \ref{stressfig}. These
stress maps confirm a number of expected results. In the absence of any loading (stage 0), we observe 
a slight increase of the stress from top to bottom, in agreement with the fact that only the weight 
of the grains themselves acts at this stage. The same pattern is observed in all directions. 
The vertical component of the stress tensor shows an increase with the loading, as well as the
radial and azimuthal components. The magnitude of the vertical component is about twice as large as the other two. 
Another finding of interest concerns the existence of the shear component $(r,z)$. It is also notable that both vertical and radial 
components of the stress tensor have larger values near the boundary of the confining cylindrical cell at the 2nd 
stage of compression. Grains at the centre experience a reduced compression, due to their ability to 
move slightly towards the sides. Lateral grains are however under stronger radial and vertical displacement 
gradients. These maps are therefore in full agreement with the displacement field measured from the tracking 
of grains for high loadings (Fig \ref{Displacement}). 

\section*{Conclusion}

We have presented in this paper the first measurements of internal stresses in a 
dry granular system resolved at the single grain scale. We have used here a simple geometry that allows us to 
calibrate and validate our measurements. In particular, 
we are now able to quantify, for the same pile, a number of characteristic features. They 
include mainly the evolution of the coordination number, the 2-point correlation function, the internal 
displacement fields, force distributions and stress fields, all as a function of loading. 
Taken together, all these measurements will enable us to better unravel the 
micromechanical behaviour of granular systems, in particular in quasistatic regimes. 
It is worth pointing out as well that by measuring the force based on the contact geometry, 
we are able to deal with non symmetric and polydisperse grains.

Although we have proven that these measurements are realistic and achievable, a number 
of improvements are still required in order to study more complex cases. 
We need, in particular, to increase both the spatial statistics (number of grains) and the resolution at 
the contact scale which will provide a better calibration of 
the force and therefore a more reliable force measurement in the pile. This will also improve our measurements 
of coordination number and other geometric quantities. Increased image resolution will also allow 
the use of stiffer grains which in turn will widen the range of suitable materials to use for the grains. These 
improvements are achievable in the near future using high resolution nano-focus CT combined with large panel detectors, 
providing at least a five fold improvement in resolution and speed. 
The contact model also needs further refinements, so that tangential forces can 
be estimated from the orientation of the contact zone with respect to the centre-to-centre 
line. 
These technical advances, combined with the analysis tools presented here, will contribute 
to the understanding of a number of open problems. Internal force distribution can now 
be studied in the bulk, reaching the low force region of the distribution. The mechanics of 
the material and in particular its mechanical stability rely on the internal organisation 
of the grain contacts and also the force network. These are quantities that can be directly 
analysed in 3D, not only statically, but also dynamically by monitoring how an applied load 
affects each individual grain contact.

The kinematic study of the material can also be qualitatively and quantitatively extended. 
Not only the local strain can be measured, but the non-affine part of the displacement 
field is also directly accessible. The latter is increasingly thought to be related with 
the bulk material properties, in particular when the system is at the onset of 
rigidity \cite{Wyart2008}. How such ideas can apply to 3D frictional granular systems remains to 
be investigated. In particular, answering such questions will require  an extension to our current 
method so that we can study grain rotation. This would allow us to test the importance of 
micropolar elasticity in the mechanics of granular systems.

\begin{acknowledgements} 

The authors wish to thank for the techincal support of ANU's X-ray tomography team in 
particular Christoph Arns and Mark Knackstedt. We thant ANU Supercomputer Facility and NCI for 
their generous allocation of computing time. We also thank Ajay Limay and Jose Maoricio 
for assisting with some of the visualisations in this article. MS acknowledges useful 
discussions with Nicolas Francois and Tomaso Aste. Financial support for this work 
through a grant from Australian Research Council, Project DP0881458, is greatefully 
acknowledged.

\end{acknowledgements}


%\bibliographystyle{spphys}
%\bibliography{paper}

\end{document}